\documentclass[11pt]{iopart}

\usepackage{iopams}
\usepackage{bm,bbm}
\usepackage{amsfonts}
\usepackage{amssymb}
\usepackage{amscd}
\usepackage{dsfont}
\usepackage{cite}
\usepackage[dvips]{graphicx}
\usepackage{euscript}
\usepackage{calc}
\usepackage{multido}
\usepackage{pst-all}

\DeclareMathAlphabet{\mathpzc}{OT1}{pzc}{m}{it}
\newtheorem{thm}{Theorem}
\newtheorem{exmp}[thm]{Example}
\newtheorem{prop}{Proposition}

\newtheorem{defn}{Definition}

\newcommand{\set}[1]{\left\{#1\right\}}

\newcommand{\C}{\mathcal{C}}
\newcommand{\chii}{\mathcal{X}}

\newcommand{\p}{$p$}
\newcommand{\ee}{ \sf e}

\newcommand{\barQ}{{\bm {\mathcal Q}}}
\newcommand{\obarQ}{{\bm \hat{{\mathcal Q}}}}
\newcommand{\barmP}{\bar{\mathcal{P}}}
\newcommand{\mP}{\mathcal{P}}
\newcommand{\bmQ}{\bm Q}
\newcommand{\hT}{\hat{T}}
\newcommand{\bmT}{\bm T}

\newcommand{\os}{\hat{s}}
\newcommand{\oQ}{\hat{Q}}
\newcommand{\oS}{\hat{S}}

\definecolor{qmuldarkblue}{rgb}{0.1490, 0.2588, 0.5882}
\definecolor{lightgrey}{rgb}{0.9,0.9,0.9}
\definecolor{lightred}{rgb}{0.9,0,0}
\definecolor{darkred}{rgb}{0.588,0,0}
\definecolor{darkdarkred}{rgb}{0.588,0.3,0.3}
\definecolor{darkdarkredtilda}{rgb}{0.588,0.25,0.25}
\definecolor{darkdarkgreen}{rgb}{0.3,0.588,0.3}

\definecolor{silver}{rgb}{0.75,0.75,0.75}
\definecolor{LightSlateGray}{rgb}{0.46484375,0.53125,0.59765625}
\definecolor{DarkSlateGray}{rgb}{0.18359375,0.30859375,0.30859375}
\definecolor{DarkSlateGray1}{rgb}{0.4,0.3,0.3}

\definecolor{backgrey}{rgb}{0.93,0.93,0.93}
\definecolor{backblue}{rgb}{0.93,0.93,1}
\definecolor{backyellow}{rgb}{1,1,0.88}

\definecolor{backred}{rgb}{1,0.9,0.9}
\definecolor{backgreen}{rgb}{0.9,1,0.9}
\definecolor{backpink}{rgb}{1,0.9,1}
\definecolor{backturquoise}{rgb}{0.9,1,1}

\begin{document}
\title[block-rectangular hierarchical matrices]{Spectral problem of  block-rectangular hierarchical matrices}
\author{ B. Gutkin, V.Al. Osipov}

\address{\it Fakult\"at f\"ur Physik, Universit\"{a}t Duisburg-Essen, 47048 Duisburg, Germany}

\ead{Vladimir.Osipov@uni-due.de, Boris.Gutkin@uni-due.de}

\date{\today}

\begin{abstract}
The spectral problem for matrices with a block-hierarchical structure is often considered in context of the theory of complex systems. In the present article, a new class of matrices with a block-rectangular non-symmetric hierarchical structure is introduced and the corresponding spectral problem is investigated. Using these results we study a model of error generation in information sequence where such  block-rectangular hierarchical matrices appear in a natural way.
\end{abstract}

\section{Introduction}
{\it Block-hierarchical matrices} are distingushed by their specific  structure of hierarchically nested growing blocks placed along the diagonal, where each (sub-)block is again a block-hierarchical  matrix itself, see Ex.\ref{ParisiMatrix_exmp}. Essential necessity in mathematical objects of such a nature arose from studying of spin-glass models at the end of the 70's~\cite{P1979, MPV1987}. In the most studied  Sherrington-Kirkpatrick model~\cite{SK1975} the block-hierarchical matrices appear as solutions of a saddle-point equation in the calculation of the free energy function by means of the replica method. The entries of these matrices are interpreted as the overlap parameter between different states of the spin-glass system which correspond to the minima of the free energy function~\cite{MPV1987}. Since the set of these minima has a hierarchically nested structure the notion of ultrametric space~\cite{RTV1986} arises in a natural way. This, in turn, is reflected in the block-hierarchical structure of the correlation matrices.
\begin{exmp}\label{ParisiMatrix_exmp}{\small A block-hierarchical matrix for $p=2$:
\begin{equation*}
\bm Q=\left(\begin{array}{c|cc}
	 \begin{array}{c|c}
	 \begin{array}{c|c}
		\begin{array}{c|c}
			q_0&q_1\\\hline
			q_1&q_0
		\end{array}        &q_2
	\\\hline
		q_2	&\begin{array}{c|c}
			q_0&q_1\\\hline
			q_1&q_0
		\end{array}
        \end{array}&q_3\\\hline
	q_3&	
		\begin{array}{c|c}
			\begin{array}{c|c}
				q_0&q_1\\\hline
				q_1&q_0
			\end{array}        &q_2
		\\\hline q_2 & 			\begin{array}{c|c}
				q_0&q_1\\\hline
				q_1&q_0
			\end{array}
		\end{array}
	\end{array}
&\dots\\\hline
	\vdots&\ddots
	\end{array}\right).
\end{equation*}
The $2\times 2$ blocks placed along the diagonal are embedded into $4\times 4$ blocks and so on. The blocks of the level-number $\gamma$, $\gamma=1,2,\dots$ have the size $2^\gamma\times 2^\gamma$.}
\end{exmp}

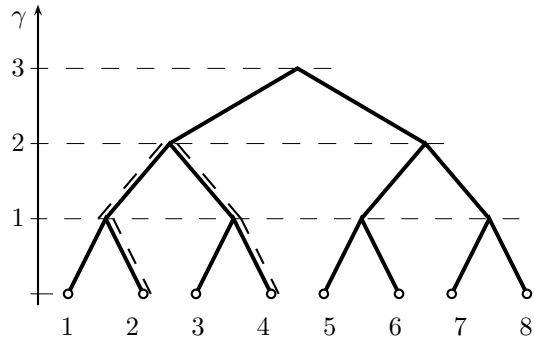
\begin{figure}
\centering
			\psset{unit=1cm,linewidth=1.5pt}
			\begin{pspicture}[showgrid=false](7.5,5)
				\multido{\nZero=1.0+1.7, \nMid=1.5+1.7, \nEnd=2.0+1.7}{4}{%
					\psline(\nZero,0.65)(\nMid,1.65)(\nEnd,0.65)
				}
				\multido{\nZero=1.5+3.4, \nMid=2.35+3.40, \nEnd=3.2+3.4}{2}{%
					\psline(\nZero,1.65)(\nMid,2.65)(\nEnd,1.65)
				}
				\psline(2.35,2.65)(4.05,3.65)(5.75,2.65)
				\multido{\nZero=1.0+1.7, \nEnd=2.0+1.7}{4}{%
					\pscircle[fillstyle=solid,linewidth=0.8pt](\nZero,0.65){2pt}
					\pscircle[fillstyle=solid,linewidth=0.8pt](\nEnd,0.65){2pt}
				}
				\psline[linestyle=dashed,dash=0.23 0.23,linewidth=0.2pt](0.5,0.65)(0.8,0.65)
				\psline[linestyle=dashed,dash=0.23 0.23,linewidth=0.2pt](0.5,1.65)(7.0,1.65)
				\psline[linestyle=dashed,dash=0.23 0.23,linewidth=0.2pt](0.5,2.65)(6.0,2.65)
				\psline[linestyle=dashed,dash=0.23 0.23,linewidth=0.2pt](0.5,3.65)(4.5,3.65)
				\psline[linewidth=0.8pt]{->}(0.6,0.5)(0.6,4.5)
				\psline[linestyle=dashed,dash=0.23 0.23,linewidth=0.8pt](1.40,1.65)(2.25,2.65)
				\psline[linestyle=dashed,dash=0.23 0.23,linewidth=0.8pt](2.45,2.65)(3.30,1.65)
				\psline[linestyle=dashed,dash=0.23 0.23,linewidth=0.8pt](1.60,1.65)(2.10,0.65)
				\psline[linestyle=dashed,dash=0.23 0.23,linewidth=0.8pt](3.30,1.65)(3.80,0.65)
{\footnotesize
				\put(0.90,0.1){$1$}
				\put(1.77,0.1){$2$}
				\put(2.64,0.1){$3$}
				\put(3.51,0.1){$4$}
				\put(4.39,0.1){$5$}
				\put(5.26,0.1){$6$}
				\put(6.13,0.1){$7$}
				\put(7.00,0.1){$8$}
				\put(0.25,4.25){$\gamma$}
				\put(0.25,1.55){$1$}
				\put(0.25,2.55){$2$}
				\put(0.25,3.55){$3$}
}
			\end{pspicture}
\caption{\small Distance graph of an ultrametric space consisting of $p^r$ elements (depicted by the numerated endpoints of the graph) with branching parameter $p=2$ and parameter $r=3$. The distance between two distinguishable elements is determined by the value $\gamma$ -- the level of hierarchy of their first common ancestor. For instance, distance between elements denoted by $2$ and $4$ is given by $\gamma=2$. }\label{Fig1}
\end{figure}

The distance, $d(x,y)$, defined on the ``ordinary'' metric spaces must satisfy the triangle inequality: $d(x,y)\le d(x,z)+d(z,y)$ where $x$, $y$ and $z$ are elements of the space. For spaces with ultrametric distance $d_u(x,y)$ some stronger relation -- the {\it strong triangle inequality} -- takes place: $d_u(x,y)\le\max\set{d_u(x,z),d_u(z,y)}$. The metric structure of these spaces can be conveniently described by means of a directed distances graph  having a tree-like structure. Elements of the ultrametric space are associated with the endpoints of the graph. The  distance between two elements $x$, $y$ is determined by the hierarchical level  $\gamma(x,y)$ of their first common ancestor on the tree, see Fig.\ref{Fig1}. The ultrametric distance $d_u(x,y)$ is a positive growing function of $\gamma$ with a supplementary constraint: $d_u(x,y)=0$ if and only if $x=y$.

Beyond the theory of spin glasses a number of works have been devoted to the investigation of non-equilibrium dynamics of various complex systems~\cite{B1997, P1999} with ultrametric structure of the free energy landscape. In these studies the block-hierarchical matrices typically  describe the  probabilities of transitions between different states of the system identified with the points of the ultrametric space. The first mathematical model of such a kind was introduced by Ogielski and Stein in~\cite{OS1985}. They considered a stochastic process which can be interpreted as ``ultrametric diffusion''. In this model the diffusion occurs on the end-points of a directed regular tree with the branching parameter $p$ (the number of outcoming branches at each vertex) and is governed by the following equation
\begin{equation}\label{Lattice_master_equation}
 \frac{d\bm f(t)}{dt}=\bm Q \bm f(t).
\end{equation}
Here  the $i$-th element of the vector $\bm f(t)$ determines the probability to find the system at the site $i$ at a given instant of time $t$. The hopping matrix $\bm Q$ of the block-hierarchical type (see Ex.\ref{ParisiMatrix_exmp}) is composed of the nested blocks of exponentially growing sizes $p^\gamma\times p^\gamma$, $\gamma=1,2,\dots,r$. The  solution of the lattice equation~(\ref{Lattice_master_equation}) reduces to the  spectral problem for the matrix $\bm Q$. Ogielski and Stein succeeded to calculate eigenvalues and eigenvectors of matrices $\bm Q$ (for $p=2$) in the limit $r\to\infty$ and found the solution of the initial-state-decay problem. Later, it was shown in~\cite{ABK1999, PS2000} that whenever the distance graph possesses a regular structure, a natural framework for the analysis of eq.(\ref{Lattice_master_equation}) is provided by the spectral theory of \p-adic pseudodifferential operators~\cite{VVZ1994}. However, for block-hierarchical matrices of a general form, where the distance graph is irregularly branched, the corresponding ultrametric space does not fit into the structure of the field of \p-adic numbers and \p-adic formalism cannot directly be applied~\cite{KhK2006}.

In this work our attention is focused on a new class of  non-symmetric block-hierarchical matrices~$\barQ$. Unlike the conventional block-hierarchical  matrices,  $\barQ$ have rectangular (rather then square) structure of  subblocks. More precisely, $p^r\times p^r$ matrix $\barQ$ is composed of $p$ identical $p^r\times p^{r-1}$ rectangular sub-matrices~$\bm O$:
\begin{equation*}
\barQ=\left(\underbrace{\bm O\dots\bm O}_p \right).
\end{equation*}
The sub-matrix $\bm O$ has the structure similar to conventional block-hierarchical matrices with the elements replaced by columns of the height $p$, as visualised in Ex.\ref{SkewParisiMatrix_exmp}. Accordingly, the nested rectangular sub-blocks in the matrix $\bm O$ are of the dimensions $p^\gamma\times p^{\gamma-1}$, $\gamma=2,\dots,r$.

\begin{exmp}\label{SkewParisiMatrix_exmp}{\small An $8\times 8$ block-rectangular hierarchical matrix~$\barQ$ with the parameters $p=2$, $r=3$:
\begin{equation*}
\fl\qquad
\barQ=\left( \begin{array}{c|c|c|c}
		\begin{array}{c|c}
			q_0& q_1\\
			q_0&q_1	\\\hline
			q_1&q_0\\
			q_1&q_0
		\end{array}        &q_2&		\begin{array}{c|c}
			q_0& q_1\\
			q_0&q_1	\\\hline
			q_1&q_0\\
			q_1&q_0
		\end{array}&q_2\\\hline
		q_2	&		\begin{array}{c|c}
			q_0& q_1\\
			q_0&q_1	\\\hline
			q_1&q_0\\
			q_1&q_0
		\end{array}&	q_2	&		\begin{array}{c|c}
			q_0& q_1\\
			q_0&q_1	\\\hline
			q_1&q_0\\
			q_1&q_0
		\end{array}
             \end{array}
\right),\qquad \bm O=\left( \begin{array}{c|c}
		\begin{array}{c|c}
			q_0& q_1\\
			q_0&q_1	\\\hline
			q_1&q_0\\
			q_1&q_0
		\end{array}        &q_2\\\hline
		q_2	&		\begin{array}{c|c}
			q_0& q_1\\
			q_0&q_1	\\\hline
			q_1&q_0\\
			q_1&q_0
		\end{array}
             \end{array}
\right).
\end{equation*}
Here $q_2$ is the block of the size $4\times 2$.}
\end{exmp}

In the present paper we aim to describe the spectral properties of the {\it block-rectangular hierarchical matrices}. As a n application, we also provide a simple error generating model, where such matrices appear in a natural way. In particular, the knowledge of the spectrum of the block-rectangular hierarchical matrices makes it possible to evaluate the average number of generated errors and the corresponding variance.

The paper is organised as follows. In the next section we introduce tensor product representation for standard block-hierarchical matrices and recall their spectral properties. In the section 3 we use this representation to introduce the class of block-rectangular hierarchical matrices and study the corresponding spectral problem. In the section 4 the error generating model is considered. The concluding remarks are given in the section 5.

\section{Block-hierarchical matrices. Tensor product representation}\label{Sec2}
The original construction of block-hierarchical matrices given in the paper~\cite{P1979} by Parisi might be described in the following way:

{\it Take $q_\gamma$, $\gamma=0,\dots,r$ be a  sequence of $r\ge1$ arbitrary numbers and take $n_\gamma$ be a sequence of positive integer numbers, such that the ratios $n_\gamma/n_{\gamma-1}$ are integers for $\gamma\ge1$. Then a non-diagonal element $\bm Q_{a\,b}$, $a\ne b$ of the $n_r\times n_r$ block-hierarchical  matrix $\bm Q$ is equal to $q_\gamma$ whenever \footnote{The brackets $ \left\lceil\cdot\right\rceil$ denote the integer part, such that for all real $x$
\begin{equation*}
\left\lceil x\right\rceil-1\leq x\leq \left\lceil x\right\rceil.
\end{equation*}
}
$$
 \left\lceil\frac{a}{n_{\gamma-1}}\right\rceil\neq\left\lceil\frac{b}{n_{\gamma-1}}\right\rceil\mbox{ and }
\left\lceil\frac{a}{n_\gamma}\right\rceil=\left\lceil\frac{b}{n_\gamma}\right\rceil.
$$
The diagonal elements $\bm Q_{a\,a}=q_0$ are allowed to be arbitrary.}

In what follows we focus on a class of regular block-hierarchical matrices, where  $n_\gamma/n_{\gamma-1}$ is the same for all $\gamma$. Below  we introduce a tensor product representation for such  matrices and reproduce the known results on their spectrum.

\subsection{Tensor product representation}\label{Sec22}
In many applications the sequence $n_\gamma$ in the above definition is chosen to be a geometric progression,
\begin{equation}\label{n_gamma}
 n_\gamma=p^{\gamma},\qquad \gamma=0,1,\dots,r.
\end{equation} 
Among a wide variety of block-hierarchical matrices this condition singles out those which have regularly growing blocks. These matrices correspond to the ultrametric spaces with regularly branching distances graphs. For instance, the Ogilski-Stein model of ultrametric diffusion~\cite{OS1985} was formulated for $p=2$, see Ex.\ref{ParisiMatrix_exmp}. In \p-adic models~\cite{ABK1999} the parameter $p$ is assumed to be a prime number as it is implied in the \p-adic formalism~\cite{VVZ1994}. In the present paper this condition is relaxed and $p$ is an arbitrary integer larger then $1$, while $r$ is a fixed integer.

The structure of block-hierarchical matrices makes it possible to present them in the form of linear combination of block-diagonal matrices. Indeed, let us introduce $p^r\times p^r$ block-diagonal matrices $\bm S_\gamma$, $\gamma=0,1,\dots,r$. Each $\bm S_\gamma$ is composed of $p^{r-\gamma}$ identical $p^\gamma\times p^\gamma$ blocks  with all entries equal to one ($\bm S_0$ is nothing but the identity matrix). It is easy to check, that a block-hierarchical matrix $\bm Q$ can then be constructed in the form of linear combination of $\bm S_\gamma$'s:
\begin{equation}\label{ParisiSum}
\bm Q =\sum_{\gamma=0}^r a_\gamma \bm S_\gamma.
\end{equation}
Each matrix $\bm S_\gamma$, in turn, can be represented as the tensor product:
\begin{equation}\label{ParisiS}
\bm S_\gamma= \underbrace{\mathds{1}\otimes\dots\otimes\mathds{1}}_{r-\gamma}\otimes \underbrace{\bm s \otimes \dots\otimes \bm s }_{\gamma},
\end{equation}
composed of $r-\gamma$ identity matrices $\mathds{1}$ and $\gamma$ projection matrices $\bm s$ of the same size $p\times p$:
\begin{equation}\label{s_def}
\bm s=\frac{1}{p}\left(\begin{array}{ccc}
 1&\dots&1 \\
 1&\dots&1 \\
 \vdots &&\vdots\\
 1&\dots&1
\end{array}\right),
\end{equation}
Here the normalisation factor $1/p$ is added to satisfy the projection property
\begin{equation}\label{matrix_s_projection_property}
\bm s^2=\bm s.
\end{equation}

The representation~(\ref{ParisiS}) together with the property~(\ref{matrix_s_projection_property}) implies that the product of matrices $\bm S_\gamma$ is
\begin{equation}\label{SCommute}
\bm S_\gamma\bm S_{\gamma'}=\bm S_{\gamma'}\bm S_\gamma=\bm S_{\max\set{\gamma,\gamma'}}.
\end{equation}
In particular, it means that both products and sums of block-hierarchical matrices are again block-hierarchical matrices.

\vspace{4pt}\noindent {\small {\bf Remark}\hspace{2pt} {\it It worth mentioning that due to the relation~(\ref{SCommute}), matrix $\bm Q$ can be cast in a product and exponential forms:
$$
\bm Q=\sum_{\gamma=0}^r a_\gamma \bm S_\gamma=e^{\sum_{\gamma}c_\gamma \bm S_\gamma}=\prod_{\gamma=0}^{r}\left(\mathds{1}+b_\gamma \bm S_\gamma \right).
$$
Equating of coefficients at different $\bm S_\gamma$ in each of the relations above results in the following connection formulae between~$a_\gamma$, $b_\gamma$, $c_\gamma$ and~$q_\gamma$:
\begin{eqnarray}\label{qAconnection}\fl
\quad q_r=a_r p^{-r},\quad q_\gamma=\sum_{\mu=\gamma}^r a_\mu p^{-\mu},\quad a_\gamma=p^\gamma(q_\gamma-q_{\gamma+1}),\quad b_\gamma=e^{c_\gamma}-1, \quad\gamma=0,\dots,r-1.
\end{eqnarray}
\begin{eqnarray*}
\fl\quad 
 a_0=1+b_0,\qquad a_\gamma=b_\gamma\prod_{\mu=0}^{\gamma-1} \left(1+b_\mu\right),
\qquad b_\gamma= a_\gamma\left/\sum_{\mu=0}^{\gamma-1} a_\mu\right. , \qquad \gamma=1,\dots,r-1.
\end{eqnarray*}
}}

\subsection{Spectral properties of $\bmQ$}\label{Sec22b}
Let us turn now our attention to the solution of the spectral problem for block-hierarchical matrices. To this end it is convenient to consider an operator formulation of the problem. To pass from matrices to operators we, first, introduce a \p-dimensional Hilbert space $h$ and the associated $p^r$-dimensional space $\mathcal{H}$ with the tensor product structure
\begin{equation}\label{SpaceH}
\mathcal{H}=\underbrace{h\otimes h\otimes\dots\otimes h}_{r}.
\end{equation}
In what follows we assume that a set of the orthonormal vectors 
\begin{equation}\label{basish_def}
 |e_j\rangle\in h,\qquad j\in\set{0,1,\dots,p-1};\qquad \langle e_i|e_j\rangle =\delta_{i,j} 
\end{equation}
forms a basis in $h$. Correspondingly, the set of the vectors
\begin{equation}\label{basis_def}
| \bm e_j>:= |e_{j_1}\rangle\otimes|e_{j_2}\rangle\otimes \dots\otimes |e_{j_r}\rangle, \qquad j_k\in \set{0,1,\dots,p-1}, 
\end{equation} 
indexed by $j=1+\sum_{\gamma=1}^rj_\gamma p^{r-\gamma}$ running from $1$ to $p^r$, provides an orthonormal basis on~$\mathcal{H}$. Let us now replace the matrix $\bm s$ in eq.(\ref{ParisiS}) with the  operator $\bm \os$ acting as a projector onto the one-dimensional subspace of $h$
\begin{eqnarray}\label{s_eigenvallue1}
\bm \os\, |e_j\rangle&=& \frac{1}{p}\sum_{i=0}^{p-1} |e_i\rangle,\qquad  j\in \set{0,1,\dots,p-1}.
\end{eqnarray}
By this definition the matrix $\bm s$, eq.~(\ref{s_def}), can be seen as the representation of $\bm \os$ in the basis~(\ref{basish_def}),  i.e. ${\bm s}_{j\;i} =\langle j| {\bm \os}|i \rangle$.

Having introduced the basic notations we define a family of operators $\mP$ reproducing the block-hierarchical matrices.
\begin{defn}\label{ClassP_def}
Let $h$ be a \p-dimensional Hilbert and let  $\mathcal{H}$ be the associated  $p^r$-dimensional Hilbert space with the structure of the tensor product~(\ref{SpaceH}). Let $\bm \os$ be the projection operator onto one-dimensional subspace of $h$, as defined by~(\ref{s_eigenvallue1}). The family $\mP$ contains all linear operators  acting on vectors $\bm\omega=\omega_{1}\otimes\omega_{2}\otimes\dots\otimes \omega_{r}\in\mathcal{H}$ according to  the rule
\begin{equation}\label{ParisiAction}
\bm \oQ \bm \omega=\sum_{\gamma=0}^r a_\gamma \bm\oS_\gamma\, \bm\omega,\qquad \bm\oS_\gamma\, \bm\omega=
\omega_{1}\otimes\dots\otimes\omega_{r-\gamma}\otimes \bm \os \omega_{r-\gamma+1}\otimes \dots\otimes \bm \os \omega_{r},
\end{equation}
where  $a_\gamma$'s are arbitrary numbers.
\end{defn}
It is easy to see that any  $\bm\oQ\in \mP$  takes a block-hierarchical form  in the basis~(\ref{basis_def}).

The tensor product form of operators $\bm\oQ\in \mP$ makes the solution of the spectral problem straightforward.
Denote $|j\rangle$, $j=1,\dots,p-1$  the  orthogonal  eigenvectors of  $\bm \os$, such that:
\begin{eqnarray*}
\bm \os\,|0\rangle&=&1\cdot|0\rangle,\\
\bm \os\, |j\rangle&=&0\cdot|j\rangle,\qquad  j \in\{1,\dots,p-1\}.\label{s_eigenvallue0}
\end{eqnarray*}
Explicitly, these eigenvectors can be expressed in the basis~(\ref{basish_def})  as
\begin{equation}\label{two_basises_connection}
|j\rangle=\frac{1}{\sqrt{p}}\sum_{k=0}^{p-1}c_k^{(j)}|e_k\rangle,\qquad  c_k^{(j)}=e^{2\pi \mathrm{i} \frac{k\,j}{p}}\qquad k,\,j\in\{0,1,\dots,p-1\}.
\end{equation}
(Note, that such a choice of the coefficients $c_k^{(j)}$ is prompted by the structure of multiplicative characters of the field of \p-adic numbers~\cite{VVZ1994}.) Then from eq.(\ref{ParisiAction}) it follows that any tensor product state
\begin{equation}\label{chi}
|j_1 j_2\dots j_r>:=|j_1\rangle\otimes|j_2\rangle\otimes \dots\otimes |j_r\rangle, \qquad j_k \in \set{0,1,\dots,p-1},
\end{equation} 
which is composed of eigenvectors of the operator $\bm \os$ is, in fact, a common eigenstate of all operators $\bm \oQ\in\mP$. Furtheremore, the above eigenstates are orthogonal to each other:
\begin{equation}\label{BasisVector_encoding}
<k_r\dots k_2k_1 |j_1 j_2\dots j_r>=\delta_{k_1,j_1}\delta_{k_2,j_2}\dots\delta_{k_r,j_r}.
\end{equation}

To find the spectrum we note that by eqs.(\ref{s_eigenvallue1}) and~(\ref{s_eigenvallue0}) each eigenvalue of $\bm\oQ$ is determined only by the number of subsequent zeroes at the end of the encoding sequence of the corresponding eigenvector. Their multiplicities, therefore, are given by the number of permutations in the remaining part of the encoding sequence $j_1 j_2\dots j_r$. This observation leads to the following spectral structure of $\bm\oQ$:
\begin{prop}\label{ParisiSpectra} Solution of the spectral problem for operators $\bm\oQ\in\mP$:
\begin{itemize}
 \item[(i)] The only nondegenerate eigenvalue corresponding to the vector $|00\dots0>$ has the form
\begin{equation}\label{ParisiSpectra1}
 \lambda^{(r)}=\sum_{\gamma=0}^r a_\gamma;
\end{equation}
 \item[(ii)] All eigenvectors
\begin{equation}\label{BasisVector}
| j_1 j_2 \dots j_{r-\mu} \underbrace{0 \dots 0}_\mu >,\qquad \mu=0,1,\dots,r-1,
\end{equation}
where $j_1, j_2, \dots, j_{r-\mu-1}$ belong to the set $\set{0,1,\dots,p-1}$ and $j_{r-\mu}\in\set{1,2,\dots,p-1}$ have one and the same eigenvalue
\begin{equation}\label{ParisiSpectra2}
\lambda^{(\mu)}=\sum_{\gamma=0}^\mu a_\gamma.
\end{equation}
The multiplicity of $\lambda^{(\mu)}$ is given by $\bm {\mathrm{ mult}}(\lambda^{(\mu)})=(p-1)p^{r-\mu-1}$.
\end{itemize}
\end{prop}

To establish connection  with the known results on the spectrum of block-hierarchical matrices one can use the relationship~(\ref{qAconnection}) between the coefficients $a_\gamma$ and $q_\gamma$. By the application of~(\ref{qAconnection}) to~(\ref{ParisiSpectra2}) we obtain  the  spectrum of block-hierarchical matrices in the same form as in~\cite{OS1985, ABK1999, PS2000}:
\begin{eqnarray}
 \lambda^{(r)}=q_0+\left(1-\frac{1}{p}\right)\sum_{\gamma=1}^r p^{\gamma}q_\gamma;\qquad\lambda^{(0)}=a_0=q_0-q_1;\\
\lambda^{(\mu)}=q_0+\left(1-\frac{1}{p}\right) \sum_{\gamma=1}^{\mu}p^\gamma q_\gamma-p^{\mu} q_{\mu+1},\qquad \mu=1,2,\dots,r-1.
\end{eqnarray}

\section{Block-rectangular hierarchical matrices} 
There are several venues to modify the block-hierarchical structure of matrices discussed above. The most straightforward way is based on  replacing of the sub-matrices entering into the tensor product representation~(\ref{ParisiS}) of $\bm S_\gamma$ with some other matrices. For instance, one can substitute the identity matrices $\mathds{1}$ with the generic diagonal matrices. This generates a class of block-hierarchical matrices with broken translational symmetry, where the block structure stays intact, while the filling of the blocks varies along the diagonal, see~\cite{KOA2005}.
Another possibility is the replacement of matrices $\bm s$. Namely, one can take a set of $p\times p$ matrices $\set{\bm s_i}_{i=1}^{r-1}$ with an arbitrary content and build up a $p^r\times p^r$ matrix in the tensor product form
\begin{equation}\label{ParisiMatrix_exmpM2}
\bm s_r\otimes \bm s_{r-1} \otimes \dots\otimes \bm s_1.
\end{equation}
In~\cite{ACNV2009} authors introduce block-hierarchical random matrices with the blocks filled up randomly by zeros and ones. Each matrix from this ensemble can be generated according to eq.(\ref{ParisiMatrix_exmpM2}) if we randomly put $\bm s_i$ to be either identity $p\times p$ matrix or upper (lower) triangular matrix of~$1$'s.

In this section we consider another modification which preserves the homogeneous block-hierarchical structure but turning the blocks themselves to be rectangular rather than square matrices as in Ex.\ref{SkewParisiMatrix_exmp}.

\subsection{Definition of block-rectangular hierarchical matrices}
At the first step we introduce an auxiliary translational operator $\bm \hT$. It is instructive to defined it through its action on the vectors of the Hilbert space $\mathcal{H}$, see eq.(\ref{SpaceH}):
\begin{defn} The translation operator $\bm \hT$ acts on vectors from $\mathcal{H}$ as follows
\begin{equation}\label{TranslationOperator_def}
    \bm \hT \omega_1\otimes  \dots \otimes\omega_{r-1}\otimes \omega_r=  \omega_r \otimes\omega_1\otimes \dots\otimes \omega_{r-1}, \qquad \omega_j\in h.
\end{equation}
\end{defn}
Note that in the basis~(\ref{basis_def}) the operator $\bm \hT$ has the following matrix representation
\newcommand{\ii}{\!\!}
\newcommand{\Ii}{\!\!\!\!}
\newcommand{\II}{\!\!\!\!}
\begin{equation}\label{TranslationOperator_MatrixRepresentation}
 \bm T=\left(\begin{array}{llllllllllllllllllllll}
\ee_1\II\II	&	&	&	&	&	&	&\ee_2\II\II&	&	&	&	&	&	&\dots	&\ee_p\II\II&	&	&	&	&		&	\\
	&\ee_1\II\II	&	&	&	&	&	&	&\ee_2\II\II&	&	&	&	&	&\dots	&	&\ee_p\II\II&	&	&	&	&	\\
	&	&\cdot\II\II&	&	&	&	&	&	&\cdot\II\II&	&	&	&	&	&	&	&\cdot\II\II&	&	&	&	\\
	&	&	&\cdot\II\II&	&	&	&	&	&	&\cdot\II\II&	&	&	&	&	&	&	&\cdot\II\II&	&	&	\\
	&	&	&	&\cdot\II\II&	&	&	&	&	&	&\cdot\II\II&	&	&	&	&	&	&	&\cdot\II\II&	&	\\
	&	&	&	&	&\ii \ee_1\II\II&	&	&	&	&	&	&\ii \ee_2\II\II&	&\dots	&	&	&	&	&	&\ii \ee_p\II\II&	\\
	&	&	&	&	&	&\ii \ee_1\II\II&	&	&	&	&	&	&\ii \ee_2	&\dots	&	&	&	&	&	&	&\ii \ee_p	\\
 \end{array}\right),
\end{equation}
with $\ee_i=(0,\dots0, 1,0 \dots 0)^T$ being the \p-dimensional vector whose only non-zero element is in the $i$-th site while all the other elements of the matrix $\bm T$ are zeroes.

Now the new family of operators $\barmP$ is introduced as products of block-hierarchical matrices and $\bm\hT$.
\begin{defn}\label{ModifiedParisMatrix_def} The family of operators $\barmP$ consists of all operators $\bm\obarQ$ of the form
$$
\bm\obarQ=\bm \hT \bm \oQ,
$$
where $\bm \oQ\in\mP$ and operator $\bm\hT$ is the translational operator~(\ref{TranslationOperator_def}).
\end{defn}
It is now straightforward to check that the representation of any operator from $\barmP$ in the basis~(\ref{basis_def}) has a block-hierarchical structure, where blocks are rectangular matrices of the dimensions $p^\gamma\times p^{\gamma+1}$, $\gamma=1,2,\dots,r$.

\subsection{Spectral properties of $\bm\obarQ$.}
To find the spectrum of operators from the family $\barmP$ we make use of a general scheme helping to reveal the spectrum of $\bm\obarQ$ by using the properties of the operator $\bm\obarQ^r$. 

For the sake of convenience, first, we introduce notations for various subsets of the eigenvectors~(\ref{chi}) of operator $\bm\oQ\in\mP$.
\begin{itemize}
\item[(i)] Let $\bm\chii^{(\mu)}$ denote the set of the vectors~(\ref{chi}) having exactly $\mu$ zeros {\it at the end} of the encoding sequences;
\item[(ii)] Let $\bm\chii^{(m;\bm\nu)}$ be the set of the vectors~(\ref{chi}) having $m$ zeros {\it in the whole} encoding sequence. The structure of zeros is defined by the vector index $\bm\nu$, which is the constraint partition of $m$:
\begin{equation}\label{partition}
\sum_{i=1}^\ell i\nu_i=m,\qquad\sum_{i=1}^\ell \nu_i\le r-m.
\end{equation} 
It is assumed here that zeros are gathered into $\ell$ clusters, separated by non-zero elements of the sequence, such that the cluster of the length $i$ enters into the sequence exactly $\nu_i$ times. Zeros standing at the end and at the beginning of the sequence are assumed to belong to one and the same cluster.
\end{itemize}

\noindent The action of the operators $\bm\oQ$ and $\bm\hT$ on the above sets of eigenvectors is given by 
\begin{eqnarray}
\bm\oQ\bm\chii^{(\mu)}&=&\lambda^{(\mu)}\bm\chii^{(\mu)}\label{Q_chi}\\
\bm\hT\bm \chii^{(\mu)}&=&\bm \chii^{(\mu-1)}, \qquad \mu\ne0,r;\qquad \bm\hT\bm \chii^{(r)}=\bm \chii^{(r)},\label{T_chi}
\end{eqnarray}
see Proposition~\ref{ParisiSpectra} and eq.(\ref{TranslationOperator_def}). From Definition~\ref{ModifiedParisMatrix_def} and eqs.~(\ref{Q_chi}), (\ref{T_chi}) it follows immediately that
\begin{eqnarray}\label{QEigenProblem}
\bm\obarQ\bm\chii^{(\mu)}=\cases{
				\lambda^{(\mu)}\bm \chii^{(\mu-1)},&$0<\mu<r$;\\
				\lambda^{(r)}\bm \chii^{(r)},&$\mu=r$.}
\end{eqnarray}
If $\mu=0$, then \
\begin{eqnarray}\label{QEigenProblem1}
\bm\obarQ\bm\chii^{(\mu)}=\lambda^{(0)}\bm \chii^{(\mu')},
\end{eqnarray}
where $\mu'$ is some arbitrary number less then $r$.

Since the operator $\bm\hT$ possesses the property $\bm\hT^r|j_1, j_2, \dots, j_r>=|j_1, j_2, \dots, j_r>$, each eigenvector of $\bm\oQ$ is simultaneously an eigenvector of $\obarQ^r$. In order to calculate the eigenvalues of $\obarQ^r$, one  needs, in addition, to take into account details of the encoding sequence of the eigenvectors:
\begin{prop}\label{Prop_Qr} All vectors $\bm\chi$ of the set $\bm\chii^{(m;\bm\nu)}$ correspond to one and the same eigenvalue of the operator $\barQ^r$:
 \begin{equation}\label{barQrEigenProblem}
 \barQ^r\bm\chi=\Lambda^{(m;\bm\nu)}\bm\chi, \qquad \bm\chi\in\bm\chii^{(m;\bm\nu)}.
\end{equation}
The eigenvalues $\Lambda^{(m;\bm\nu)}$ are expressed in terms of $\lambda^{(\mu)}$ in the following form:
\begin{equation}\label{barQrEigenvalues}
\Lambda^{(m;\bm\nu)}=\cases{	\left( \lambda^{(m)}\right)^r,&$m=0,r$;\\
			\left( \lambda^{(0)}\right)^{r-m}\prod_{i=1}^\ell\left(\lambda^{(i)}\lambda^{(i-1)}\dots\lambda^{(1)} \right)^{\nu_i},&$0<m<r$.}
\end{equation}
The corresponding multiplicities are given by
\begin{equation}\label{barQrEigenvaluesMultiplicity}
\fl\qquad \bm {\mathrm {mult}}\left(\Lambda^{(m;\bm\nu)} \right) =\cases{(p-1)^r,	&$m=0$;\\
			r(p-1)^{r-m}\frac{(r-m-1)!}{(r-m-\sum_i\nu_i)!\prod_i \nu_i!},&$0<m<r$;\\
			1,&$m=r$.}
\end{equation}
\end{prop}
The particular form of the eigenvalues follows from eqs.(\ref{QEigenProblem}),~(\ref{QEigenProblem1}) with  the multiplicities  determined by the number of vectors forming the set $\bm \chii^{(m;\bm\nu)}$. The latter combinatorial problem can be also thought as the m\'enage problem  for counting of the number of different ways in which it is possible to set $\ell$ zeros' clusters over $r-m$ places ($m<r$, $\ell=\sum_i\nu_i\le r-m$). The number of choices of non-zero elements in the sequence is accounted by the multiplier $(p-1)^{r-m}$, while the term $r$ counts the number of vectors obtained by all possible cyclic rotations of the sequence.

Having found the spectrum of~$\bm\obarQ^r$ we can now solve the spectral problem for the operator $\bm\obarQ$ itself,
\begin{equation}\label{barQEigenProblem}
 \bm\obarQ \bm\psi=\Lambda\bm\psi.
\end{equation}
Given an eigenvector $\bm\chi\in{\bm\chii}^{(m;\bm\nu)}$ of $\bm\obarQ^r$ one can construct the set of eigenvectors $\bm\psi_l$ of the operator~$\bm\obarQ$ in the following way
\begin{equation}\label{psi_sol}
\bm \psi_l=\left( \sum_{k=0}^{r-1} \delta_l^k \bm\obarQ^k\right) \bm\chi, \qquad  \delta_l=\left(\Lambda^{(m;\bm\nu)}\right)^{1/r} e^{\frac{2\pi \mathrm{i}l}{r}},\qquad l=1,\dots r,
\end{equation}
the $\ell$-th eigenvector corresponds to the eigenvalue $\Lambda=1/\delta_l$. Note, that if the parameter $r$ is not a prime number then some of the above vectors $\bm \psi_l$ might vanish. It happens when the primary vector $\bm\chi$ has the ``period'' $d$ less then $r$, i.e. the following relation $\bm\hT^d{\bm \chi}={\bm \chi}$ takes place at $d<r$. For these vectors the formula (\ref{psi_sol}) generates only $d$  eigenvectors of $\bm\obarQ$ but not $r$. To avoid a combersome formulation of the spectral theorem below we assume that the parameter~$r$ is a prime number. In this case the only possible values of the period~$d$ are either~$1$ or~$r$. Observing that two eigenvectors $\bm\chi$ and $\bm\chi'$ generate the same set of eigenvectors of~$\bm\obarQ$ if and only if they are related by $\bm\chi={\bm T}^k \bm\chi'$ for some $k$, we obtain  the following spectral structure of the operators from $\barmP$:

\begin{prop} Let $r$ be a prime number, then operators $\bm\obarQ\in\barmP$ have the following spectral data
\begin{itemize}
 \item[(i)] Case $m=r$. The only nondegenerate eigenvalue is $\lambda^{(r)}$. It corresponds to the eigenvector~$|00\dots 0>$;
 \item[(ii)] Case $m=1,2,\dots,r-1$. The eigenvalues
$$
\Lambda^{(j,m;\bm\nu)}=\left(\Lambda^{(m;\bm\nu)}\right)^{1/r}\cdot e^{2\pi\mathrm{i} \frac{j}{r}},\quad \bm {\mathrm{ mult}}\left(\Lambda^{(j,m;\bm\nu)}\right) =\frac{1}{r}\;\bm {\mathrm{ mult}}\left(\Lambda^{(m;\bm\nu)}\right), \quad j= 0,1,\dots,r-1
$$
correspond to the set of eigenvectors $\bm \psi^{(j,m;\bm\nu)}$ having the form
\begin{equation}\label{barQrEigenvectors}
\bm \psi^{(j,m;\bm\nu)}=\frac{1}{\sqrt{C}}\sum_{k=0}^{r-1}\frac{e^{-2\pi\mathrm{i} \frac{j(k-1)}{r}}}{\left(\Lambda^{(m;\bm\nu)}\right)^{\frac{k-1}{r}}} \, \bm\obarQ^k \bm\chi, \qquad \bm\chi \in {\bm\chii}^{(m;\bm\nu)},
\end{equation}
where $C$ is a constant fixed by the normalisation condition;
 \item[(iii)]  Case $m=0$. The eigenvalue $$
\Lambda^{(0,0;\bm 0)}=\lambda^{(0)} ,\qquad \bm {\mathrm{ mult}}\left(\Lambda^{(0,0;\bm 0)}\right)=\frac{(p-1)^r-(p-1)}{r} +p-1, 
$$
corresponds to the eigenvectors obtained  by application of~(\ref{barQrEigenvectors}) to the vectors $\bm\chi$ from the set  ${\bm\chii}^{(0;\bm 0)}$.
\end{itemize}
\end{prop}

\section{Errors generation model in information sequences}
In this section we provide a schematic model of error generation in information sequences where the block-rectangular hierarchical matrices appear in a natural way.
\begin{figure}[t]
\parbox{0.4 \textwidth}{
(a)
\vspace{5pt}
\begin{center}
			\psset{unit=1cm,linewidth=1.pt}
			\begin{pspicture}[showgrid=false](5,5)
			\pscircle(2,3){2}
			\pswedge(2,3){2}{-90}{-70}
			\pswedge(2,3){2}{-50}{-30}
			\pswedge(2,3){2}{-10}{10}
			\pscircle[fillstyle=solid](2,3){1.4}
			\psarc[linecolor=darkdarkred]{<-}(2,3){0.6}{-60}{190}
			\psarc[linestyle=dotted,linewidth=3.pt,dotsep=3pt](2,3){1.7}{20}{40}
			\psarc[linestyle=dotted,linewidth=3.pt,dotsep=3pt](2,3){1.7}{-120}{-100}
{\tiny
			\put(2.26,1.28){\color{qmuldarkblue} $\mathbf 0$}
			\put(2.80,1.47){\color{qmuldarkblue} $\mathbf 1$}
			\put(3.24,1.84){\color{qmuldarkblue} $\mathbf 1$}
			\put(3.53,2.35){\color{qmuldarkblue} $\mathbf 0$}
			\put(3.64,2.90){\color{qmuldarkblue} $\mathbf 1$}
}
			\psline[linecolor=darkdarkred,linewidth=1.pt]{->}(2,0.4)(2,0.95)
			\psarc[linecolor=lightred,arcsepA=1.pt,linewidth=0.8pt]{<->}(2,3){2.35}{-90}{22}
{\small
			\put(1.68,0.35){\color{qmuldarkblue} $S$}
			\put(3.5,1.0){\parbox{1cm}{error}}
			\put(3.5,0.7){\parbox{1cm}{zone}}
}
			\end{pspicture}
\end{center}
}\hfill
\parbox[c]{0.55\textwidth}{(b)
\vspace{5pt}
\begin{center}
\includegraphics[height=0.3\textwidth,width=0.5\textwidth]{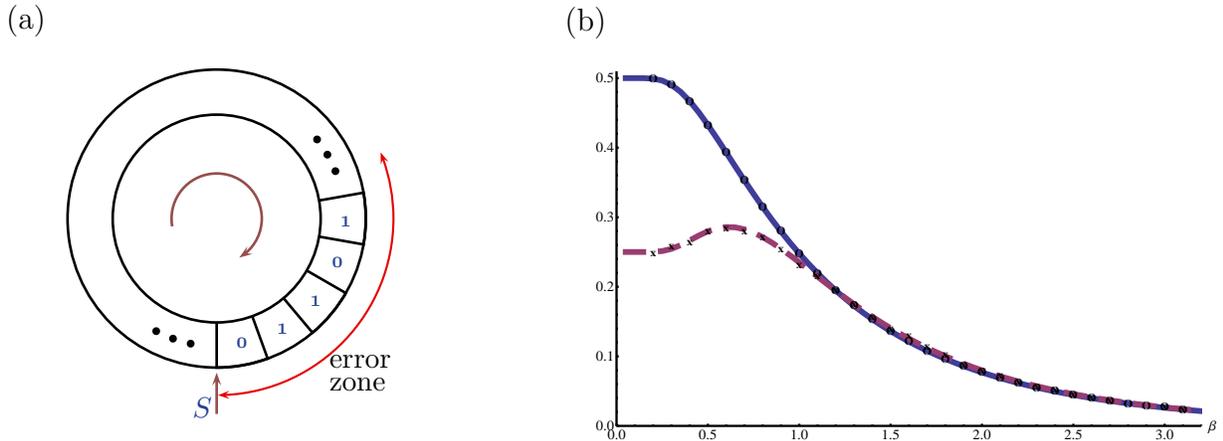}
\end{center}
}
\caption{\label{fig3} \small (a) The scheme of the model of error generation in information sequences. The information carrier rotating clockwise is under the influence of external noise. It changes the content of several consecutive cells starting from the position $S$. The level of noise defines a number of cells affected at one time-step.

(b) The mean value of errors (solid curve) accumulated in the information sequence and its variance (dashed curve) after the full cycle ($r=100$) plotted as a function of the inverse temperature $\beta$ for the model of the Boltzmann noise. The data of the numeric experiment for the same model of noise are depicted by circles and crosses, respectively.}
\end{figure}

\subsection{Formulation of the model}
The model is formulated in the following way. We assume that the information carrier (``disk'') has a circular shape and consists of $r$ cells $\C_j$, $j=1,\dots r$. Each of them contains a digit from the set $\set{0,\dots,p-1}$.  At the discreet moments of time $t=0,1,\dots$ disc rotates one step clockwise. The information on the carrier can be changed between rotations of the disk due to an external noise which generates errors in a number of neighbouring cells. More specifically:
\begin{itemize}
 \item[--] 
The noise affects the cells placed ``rightwards'' to some fixed point, $S$ (see Fig.~\ref{fig3}). At the initial moment of time $t=0$, the cell $\C_0$ is located at the point $S$. Disk rotates such that at the moment of time $t=\ell\le r$ the point $S$ points to the cell $\C_\ell$.
\item[--]
The intensity level of the noise determines the number $\gamma\in \set{0,\dots,r}$ of consecutive cells $\C_\ell,\dots,\C_{\ell+\gamma}$, whose information content changes randomly at the $\ell$-th time step. The probability to affect $\gamma$ consecutive cells is given by $a_\gamma$. The content of each affected cell can be replaced by any symbol from the set $\set{0,\dots,p-1}$ with equal probability.
\end{itemize}

In the probabilistic language each state of the disc can be described by a vector $\bm w$ from the Hilbert space $\mathcal{H}$ (see eq.(\ref{basis_def})):
\begin{equation}\label{vector_w_def}
\fl\qquad \bm w=\sum_jw_j|\bm e_j>, \qquad |\bm e_j>=|e_{j_1}\rangle\otimes|e_{j_2}\rangle\otimes \dots\otimes |e_{j_r}\rangle, \qquad j_k\in \{0,1,\dots,p-1\},
\end{equation}
where $w_j$ is the probability to find the disc with the content given by the sequence of symbols $j_1 j_2\dots j_r$ corresponding to $j$. The initial state of the system is described by the vector $\bm w(0)=|\bm e_{j'}>$ for some $j'$. After a time~$t$ the system evolves into another state
\begin{equation}\label{vector_w_of_t}
 \bm w(t)=\sum_jw_j(t)|\bm e_j>,
\end{equation}
where the coefficients $w_j(t)$ define probabilities to find the disc having informational content~$j_1 j_2\dots j_r$.

The time evolution of the system can be naturally separated into two processes: rotation of the disc and change of the cells content due to the noise. It is easy to see that the noise effect on $\gamma$ consecutive cells is described by the operator $\bm\oS_\gamma$ taken with the probability weight $a_\gamma$. The particular form of the operator $\bm\os$ in the definition of $\bm\oS_\gamma$, eq.(\ref{ParisiAction}), ensures the equidistribution of the resulting content of the affected cells. The time evolution corresponding to the  rotation of the disc is provided by the action of the operator~$\bm\hT$. Thus, the result of $\ell$ steps time evolution is described by the operator
\begin{equation}
\prod_{i=1}^\ell\bm\hT \sum_{\gamma_i=0}^r a_{\gamma_i}  \bm\oS_{\gamma_i} = (\bm\hT\bm\oQ)^\ell=\bm\obarQ^\ell
\end{equation}
acting on the initial vector $\bm w(0)$. Note, that the probabilistic interpretation of the coefficients $a_\gamma$'s implies that $\sum_{\gamma=0}^ra_\gamma=1$. By eq.(\ref{ParisiSpectra1}) this constraint is equivalent to $\lambda^{(r)}=1$.

\subsection{Estimation of losses} In this section we calculate the mean value of errors $<k>$ accumulated in the sequence after one full turn of the disk and the corresponding variance $<(k-<k>)^2>$. For the sake of simplicity of exposition the parameter $p$ is chosen below to be equal $2$. To perform the calculation we use the generation function $R_\ell(\alpha)$ for the moments of the number of errors accumulated after $\ell$ steps of time evolution. The sought moments can be calculated through the formulae:
\begin{eqnarray}\label{moment1}
\qquad< k >_{t=\ell} &=& \partial_\alpha R_\ell(\alpha)|_{\alpha=0};\\\label{moment2}
<(k-<k>)^2>_{t=\ell}&=&\partial^2_\alpha R_\ell(\alpha)|_{\alpha=0}-< k >_{t=\ell}^2.
\end{eqnarray}
To construct the generation function we introduce an auxiliary operator $\hat{\bm E}$ of the form:
\begin{eqnarray}\label{operator_E}
\hat{\bm E}=(\mathds{1}+e^\alpha\;\mathds{P})\otimes (\mathds{1}+e^\alpha\;\mathds{P})\otimes\dots     \otimes (\mathds{1}+e^\alpha\;\mathds{P});\\
\mathds{1}=|e_0\rangle\langle e_0|+|e_1\rangle\langle e_1|,\qquad\mathds{P}=|e_0\rangle\langle e_1|+|e_1\rangle\langle e_0|,\nonumber
\end{eqnarray}
where $|e_0\rangle,|e_1\rangle\in h$ and $\alpha$ is a parameter. In the basis~(\ref{basis_def}) this operator has the following matrix elements
\begin{equation}\label{E_basic_def}
<\bm e_j|\hat{\bm E}|\bm e_{j'}>=e^{k\alpha},
\end{equation}
with $k$ being the number of different symbols (compared pairwise) in the sequences $j$ and~$j'$. Due to the property~(\ref{E_basic_def})  the moment generation function can be written as
\begin{equation}\label{error_moments_generator}
  R_\ell(\alpha)=<\bm e_{j_0}|\bm\hT^{-\ell}\hat{\bm E}\bm\obarQ^\ell|\bm e_{j_0}>.
\end{equation}
where $j_0$ encodes the initial information sequence. We apply this formula to estimate the mean value of errors generated after the full cycle of disk, i.e. when $\ell=r$. Under the latter assumption $\bmT^{-r}\equiv \mathds{1}$, so that
$$
 R_r(\alpha)=2^{-r}\tr \hat{\bm E}\bm\obarQ^r.
$$
Here we also used the fact of independence of $ R_r(\alpha)$ on the initial state $j_0$. It is convenient to calculate the above trace in the basis~(\ref{chi}) where $\bm\obarQ^r$ and $\hat{\bm E}$ take the diagonal form. Therefore, we arrive to the general expression for the moment generation function
\begin{eqnarray}
\fl  R_r (\alpha)= 2^{-r}\sum_{m,\bm\nu}\Lambda^{(m;{\bm\nu})}\;\tr_{{\bm\chi}^{(m;{\bm\nu})}}\hat{\bm E}  =\underbrace{2^{-r}(\lambda^{(r)})^r(1+e^\alpha)^r}_{m=0}+ \underbrace{2^{-r}(\lambda^{(0)})^r(1-e^\alpha)^r}_{m=r}
\nonumber\\ \fl\qquad\qquad\qquad\qquad\qquad+2^{-r}\sum_{m=1}^{r-1}(1+e^\alpha)^m(1-e^\alpha)^{r-m}\sum_{\bm\nu}\Lambda^{(m;{\bm\nu})}\mathbf{mult}(\Lambda^{(m;{\bm\nu})})
\end{eqnarray}
By the formulae~(\ref{moment1}),~(\ref{moment2}) taken at $\ell=r$ we obtain
\begin{eqnarray}\label{mu}
\qquad\qquad<k >&=& \frac{r}{2}\bigg(1-g(r-1) \bigg) ;\\
<(k-<k>)^2>&=&\frac{r}{4}\left(1-r g(r-1)^2+\sum_{j=0}^{r-2}g(r-2-j)g(j)\right),\label{sigma}
\end{eqnarray}
where we used the fact that $\lambda^{(r)}=1$ and introduced the notation $g(\ell)=\prod_{i=0}^{\ell} \lambda^{(i)}$. 

It is interesting to investigate behaviour of the moments in the case of infinitely long sequences. Taking the limit $r\to \infty$ in the expressions~(\ref{mu}),~(\ref{sigma}) and assuming that $\lim_{r\to\infty}g(r)=g_\infty$ exists and the sum $G=\sum_{j=0}^\infty(g_\infty-g(j))$ converges we obtain
\begin{eqnarray}\label{mu1}
<k >&=& \frac{r}{2}\bigg(1-g_\infty \bigg) +\mathcal{O}(r^0);\\
<(k-<k>)^2>&=&\frac{r}{4}\left(1- g_\infty^2+2g_\infty G\right)+\mathcal{O}(r^0).\label{sigma1}
\end{eqnarray}
Note, that both average and variance divided by the total number of cells become finite functions in the large-$r$ limit.

As an example, consider the case of Boltzmann noise where we take the probabilities $a_\gamma$ to be proportional to $e^{-\beta\gamma}$ with parameter $\beta$ playing a role of the inverse temperature. From the normalising condition $\sum_{\gamma=0}^ra_\gamma=1$ it follows that
$$
a_\gamma=\frac{1-e^{-(r+1)\beta}}{1-e^{-\beta}}e^{-\beta\gamma}.
$$
By the formula~(\ref{ParisiSpectra2}) the eigenvalues can be expressed explicitly as
\begin{equation}\label{lambda_temperature}
\qquad\qquad\qquad\lambda^{(\gamma)}=\frac{1-e^{-(\gamma+1)\beta}}{1-e^{-(r+1)\beta}},
\end{equation}
so that
$$
g(\ell)=\frac{\prod_{\gamma=1}^{\ell+1}(1-e^{-\gamma\beta})}{(1-e^{-(r+1)\beta})^{\ell+1}},\qquad g_\infty=\prod_{\gamma=1}^{\infty}(1-e^{-\gamma\beta}).
$$
This can be substituted into eqs.~(\ref{mu1}),~(\ref{sigma1}). The resulting average and variance are plotted as functions of $\beta$ and compared with the data of the numerical experiment on Fig.\ref{fig3} (b).

One can see that at high temperatures (low $\beta$) the mean number of errors approaches the half, while the variance -- the quarter of the mean value of the full number of elements in the sequence. In this regime the noise affects almost every cell in the sequence leading to the Binomial distribution of the number of errors (the probability that exactly $k$ cells change their content is $2^{-r}\left(r\atop k\right)$). This explanes the relationship between the first and the second moments observed at high temperatures.

In the opposite regime of very low temperatures (large $\beta$) the randomising events caused by the noise become rare and mostly affect one cell only. This results in the Poisson process characterised by the equality between the mean value and the variance clearly observed on the plot (see the right tails of the plot, Fig.\ref{fig3} (b)). Indeed, at low temperatures the quantity $g(k)$ can be approximated by
$$
\beta\gg1:\qquad g_\infty=g(k)= 1-e^{-\beta}+\mathcal{O}(e^{-2\beta}),
$$
which, in turn, yields
$$
<k >=<(k-<k>)^2>= \frac{re^{-\beta}}{2}\bigg(1+\mathcal{O}(e^{-\beta}) \bigg).
$$

In the intermediate regime ($\beta\sim1$) some of the randomising events start to affect clusters of cells rather then individual cells. This leads to some decrease of the variance in comparison with the one obtained from the Poisson process. An interesting feature is that at some point the variance reaches its maximal value and then decreases at higher temperatures. Qualitatively, this can be explained in the following way. At large $\beta$ the variance grows with the temperature since the total number of the randomising events also increases. On the other hand, for high temperatures the large-scale randomising events effectively ``swallow'' the smaller ones, actually, reducing the relevant number of randomising events. Eventually, this process becomes dominant at very high temperatures and leads to decreasing of the variance.

\section{Discussion}
In the present article a new family of block-rectangular (non-symmetric) hierarchical matrices $\barQ$ was introduced. The latter were utilised to construct the model of error generation in information sequences. For this model the mean number of errors and variance were calculated and estimated for a particular realisation of the noise.

It should be emphasised that the evolution generated by $\barQ$ can be considered as a composition of two kinds of processes: the stochastic  ultrametric diffusion described by the operator $\bm \oQ$ and deterministic dynamics induced by $\bm \hT$. Since the operator $\bm \hT$ is nothing but the shift map on symbolic sequences, the resulting dynamics has a chaotic character~\cite{PZK2001, G2010}. In particular, it is easy to see that due to the presence of $\bm \hT$ the fully ergodig state (completely randomised sequence) is achieved much faster then in the pure ultrametric diffusion. The time necessary for the equilibration is determined by the gap in the spectrum between~$1$ and the next largest eigenvalue of the evolution operator. In the framework of our noise model this quantity can be estimated for the pure ultrametric diffusion as
$$
\lambda^{(r)}=1,\qquad \lambda^{(r-1)}\sim1-(1-e^{-\beta})e^{-r \beta},
$$
see eq.~(\ref{lambda_temperature}). The characteristic equilibration time is $\tau_{\mbox{\scriptsize eq}}=-1/\ln(\lambda^{(r-1)})\sim e^{r \beta}(1-e^{-\beta})^{-1}$ and thus becomes huge in the large-$r$ limit. For the operator $\bm\obarQ$ the closest eigenvalue to $1$ is equal to $g(r-1)^{1/r}$. When $r\to\infty$ it can be estimated as
$$
e^{-\frac{\pi^2}{6r \beta}}\le g(r-1)^{1/r}\le ( 1-e^{-\beta})^{1/r}.
$$
The equilibration time therefore grows linearly with $r$ and bounded from above by $\tilde{\tau}_{\mbox{\scriptsize eq}}\lesssim r e^\beta$.

Generally speaking, the time evolution generated by $\bm\obarQ$ can be seen as a diffusion in the ultrametric space whose structure changes dynamically due to some external factors. In the model of error generations considered above this factor was rotation of the disk. We believe that similar constructions might arise in other systems having (quasi-)ultrametric state space, e.g. spin-glasses,  macromolecules, polymers etc., subjected to time-dependent forces.

\section*{Acknowledgements}
We thank Dr. S.K. Nechaev for stimulating our interest in the subject and Prof. Dr. H.-J. Sommers for valuable discussions. This work was supported by the Sonderforschungsbereich Transregio~12.

\end{document}